\documentclass[10pt,conference]{IEEEtran}
\IEEEoverridecommandlockouts
\usepackage{cite}
\usepackage{amsmath,amssymb,amsfonts}
\usepackage{graphicx}
\usepackage{textcomp}
\usepackage{anyfontsize}
\usepackage{xcolor}
\def\BibTeX{{\rm B\kern-.05em{\sc i\kern-.025em b}\kern-.08em
    T\kern-.1667em\lower.7ex\hbox{E}\kern-.125emX}}

\usepackage[frozencache=true,cachedir=minted-cache]{minted}
\AtBeginEnvironment{minted}{%
  }
\usepackage{hyperref}
\usepackage{caption}
\usepackage{subcaption}
\usepackage{booktabs}
\usepackage{color, colortbl}
\definecolor{Gray}{gray}{0.9}
\usepackage{algorithm}
\usepackage{algpseudocode}

\usepackage{tcolorbox}
\usepackage{etoolbox}

\begin{document}

\title{Polyglot Code Smell Detection for Infrastructure as Code with GLITCH}

\author{\IEEEauthorblockN{Nuno Saavedra\IEEEauthorrefmark{1}, João Gonçalves\IEEEauthorrefmark{1}, Miguel Henriques\IEEEauthorrefmark{1}, João F. Ferreira\IEEEauthorrefmark{1} and Alexandra Mendes\IEEEauthorrefmark{2}}
\IEEEauthorblockA{\IEEEauthorrefmark{1}INESC-ID and IST, University of Lisbon, Lisbon, Portugal}
\IEEEauthorblockA{\IEEEauthorrefmark{2}HASLab / INESC TEC \& Faculty of Engineering, University of Porto, Porto, Portugal}}
%
\def\GLITCH{GLITCH}
\def\numsecsmells{nine}
\def\numdesimpsmells{nine}

\newcommand\TODO[1]{%
  \ifx&#1&%
    \textcolor{red}{TODO!}
  \else
    \begin{flushleft}
      \color{red}%
      TODO: #1%
    \end{flushleft}
  \fi
  \typeout{*** TODO: #1}%
}

\def\review#1{#1}
\def\reviewjffafm#1{#1}

\newcommand\unindentedparagraph[1]{\par\smallskip\noindent\textit{\textbf{#1}:}}

\maketitle

\begin{abstract}
This paper presents \GLITCH, a new technology-agnostic framework that enables automated polyglot code smell detection for Infrastructure as Code scripts. \GLITCH\ uses an intermediate representation on which different code smell detectors can be defined. It currently supports the detection of \numsecsmells\ security smells and \numdesimpsmells\ design \& implementation smells in scripts written in Ansible, Chef, Docker, Puppet, or Terraform.
Studies conducted with \GLITCH\ not only show that \GLITCH\ can reduce the effort of writing code smell analyses for multiple IaC technologies, but also that it has higher precision and recall than current state-of-the-art tools. 
A video describing and demonstrating \GLITCH\ is available at: \url{https://youtu.be/E4RhCcZjWbk}.
\end{abstract}




\section{Introduction}
\label{sec:introduction}
Infrastructure as Code (IaC) is the process of managing IT infrastructure via programmable configuration files (also called IaC scripts). 
In recent years, several tools for detecting code smells in IaC scripts have been proposed~\cite{sharma2016does,schwarz2018code,rahman2019seven,rahman2021security,reis2022leveraging,opdebeeck2023control}.
These tools are very valuable, since they cover a wide range of code smells and several major IaC technologies. However, their implementations are separate and involve substantial duplication. If one wishes to implement the detection of a new smell, one has to develop a different implementation for each of the IaC technologies supported. 
Consequently, it is often the case that the detection of code
smells is inconsistent for different IaC technologies. For example, \review{Figure~\ref{fig:no-smell-ex1} presents a line of code with a comment taken from the project \textit{puppet-foreman} by The Foreman\footnote{\url{https://github.com/theforeman/puppet-foreman/blob/1d09876d7838bcd133add6266f4ba19b936ccb6c/manifests/init.pp\#L57}}.
For this example, Schwarz et al.'s tool~\cite{schwarz2018code} detects the \emph{Long Statement} smell because the line has exactly 140 characters, and the tool reports the smell for lines with 140 characters or more. However, if we use the tool Puppeteer~\cite{sharma2016does} to analyze the same line in Puppet, the smell will not be detected since Puppeteer only detects the \emph{Long Statement} smell for lines with more than 140 characters.
}
\reviewjffafm{Even though this example might be considered a minor problem, it shows that having separate implementations for code smell analysis can easily lead to inconsistent code smell detection. Ensuring consistency is particularly important for projects that use more than one IaC technology.}



\reviewjffafm{To address this problem,}
we present \GLITCH, a technology-agnostic framework that enables automated polyglot smell detection by transforming IaC scripts into an intermediate representation, on which different code smell detectors can be defined. \GLITCH\ currently supports the detection of \numsecsmells\ security smells and \numdesimpsmells\ design \& implementation smells in scripts written in Ansible, Chef, Docker, Puppet,  or Terraform. A previous study compared \GLITCH\ with state-of-the-art security smell detectors~\cite{saavedra2022glitch}. 
In this paper we introduce an extended version of \GLITCH\ that supports 9 previously unreported design \& implementation code smell detectors and that extends the original tool with support for Docker and Terraform. We also present preliminary results on the detection of design \& implementation smells for Ansible, Chef, and Puppet. 

The envisioned users of \GLITCH\ are \reviewjffafm{DevOps engineers and} system administrators who have to develop or maintain IaC scripts. \GLITCH\ is particularly helpful in environments where multiple IaC technologies are being used, which happens in many organizations. Moreover, since we created and make available three large datasets containing 196,756 IaC scripts (with a total of 12,281,383 LOC), and three oracle datasets for security smells (one for each IaC technology supported by \GLITCH), we argue that \GLITCH\ can also be used by researchers interested in software quality of IaC scripts. 

\GLITCH\ is open-source and is available online at: {\url{https://github.com/sr-lab/GLITCH}}. A Docker container to replicate the study on design \& implementation smells presented in this paper is available at: {\url{https://doi.org/10.6084/m9.figshare.21407058.v1}}.

\begin{figure*}
    \centering
    \fontsize{6}{7}
    \begin{tcolorbox}[left=0pt,right=0pt,top=4pt,bottom=4pt]
    \begin{minted}[]{text}
# $unattended_url::               URL hosts will retrieve templates from during build (normally http as many installers don't support https)
    \end{minted}
    \end{tcolorbox}
    \caption{Line of code taken from puppet-foreman by The Foreman. Issues with state-of-the-art tools: Schwarz et al.'s tool~\cite{schwarz2018code} reports the smell ``Long Statement'' since the line has exactly 140 characters. Puppeteer~\cite{sharma2016does} does not report the smell since its rule only considers lines with more than 140 characters.}
    \label{fig:no-smell-ex1}
\end{figure*}


\section{\GLITCH}
\begin{figure}
    \centering
    \includegraphics[width=1\columnwidth]{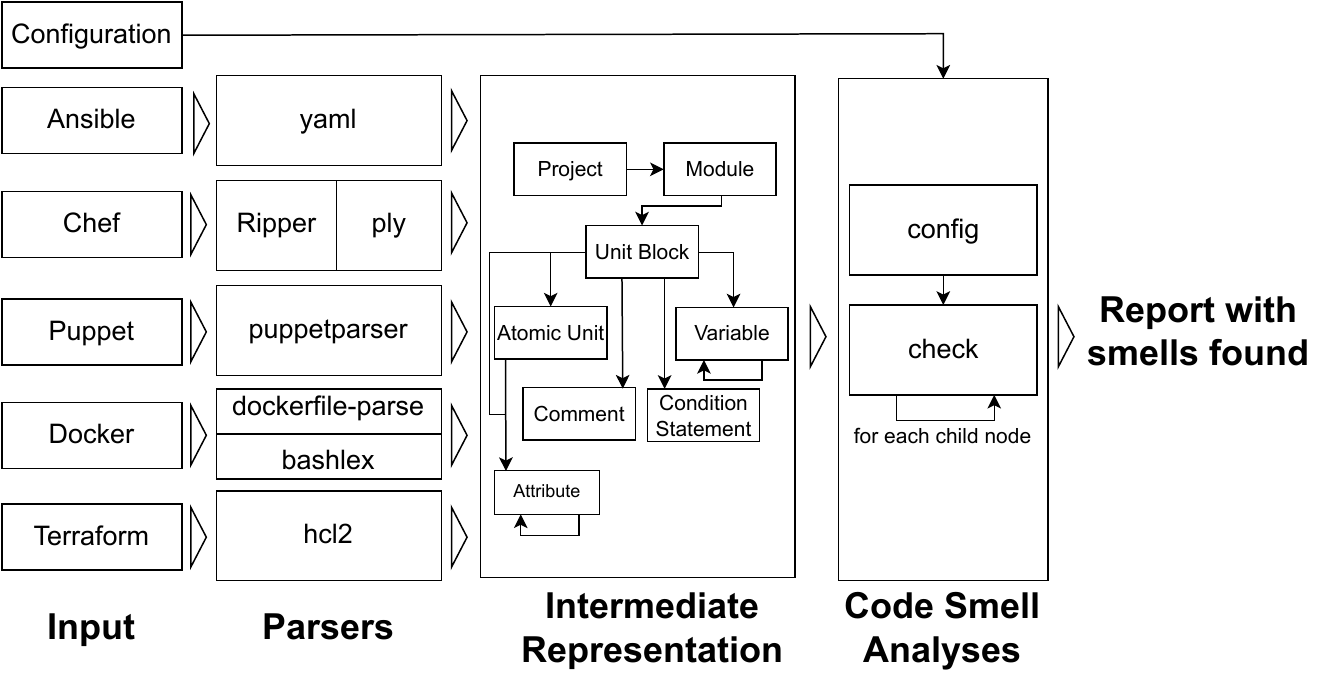}
    \caption{\GLITCH's architecture overview.}%
    \label{fig:architecture}
\vspace{-1em}
\end{figure}

This section describes \GLITCH, providing an overview of the intermediate language and the methodology for smell detection. It also provides information about \GLITCH's implementation and how it can be used. Figure~\ref{fig:architecture} shows an overview of \GLITCH's architecture.

\subsection{Intermediate Language}

\begin{figure*}
    \begin{minipage}[c][5.6cm][t]{.5\textwidth}
    \vspace{1.5em}
    \scriptsize
    \begin{minted}[]{python}
        # Hive metastore MySQL database need a (...)
        exec { 'hive_mysql_create_database':
          command => "/usr/bin/mysql (...)",
          unless => "/usr/bin/mysql (...)",
          user => 'root',
        }
    \end{minted}
    \subcaption{Part of a Puppet script from a CDH module by Wikimedia\footnotemark.}
    \label{fig:puppet-script}
    \vspace{1.5em}
    \scriptsize
    \begin{minted}[]{ruby}
        # Hive metastore MySQL database need a (...)
        execute 'hive_mysql_create_database' do
            command "/usr/bin/mysql (...)"
            not_if "/usr/bin/mysql (...)"
            user 'root'
        end
    \end{minted}
    \subcaption{Script above written in Chef.}
    \label{fig:chef-script}
    \end{minipage}
    \begin{minipage}[c][5.6cm][t]{.5\textwidth}
        \centering
        \includegraphics[width=0.9\textwidth]{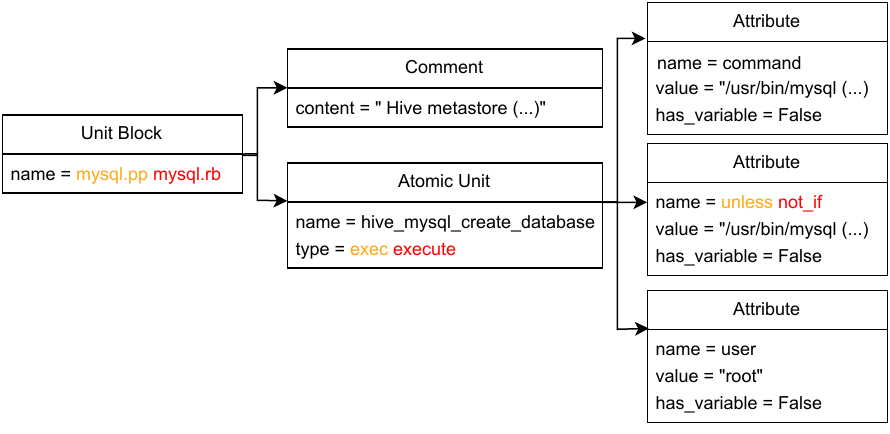}
        \subcaption{Graph-based representation of the scripts on the left using our intermediate representation. In black and orange: the representation of the Puppet script from Figure \ref{fig:puppet-script}. In black and red: the representation of the Chef script from Figure \ref{fig:chef-script}.}
    \end{minipage}
    \caption{Translation of scripts to GLITCH's intermediate representation.}%
    \label{fig:parsing}
\end{figure*}

The intermediate representation used by \GLITCH\ captures similar concepts from different IaC technologies. The representation uses a hierarchical structure, as shown in Figure~\ref{fig:architecture}. Projects represent a generic folder that may contain several modules and unit blocks. \emph{Modules} are the top component from each structure and they agglomerate the scripts necessary to execute a specific functionality (corresponding to roles in Ansible, cookbooks in Chef, and modules in Puppet and Terraform). 
\emph{Unit Blocks} correspond to the IaC scripts themselves or to a group of atomic units (corresponding to playbooks in Ansible, recipes in Chef, scripts and build stages in Docker, classes in Puppet, and configuration files in Terraform). Finally, \emph{Atomic Units} define the system components we want to change and the actions we want to perform on them (corresponding to tasks in Ansible, bash commands in Docker, and resources in Chef, Puppet, and Terraform). 
\review{Figure~\ref{fig:parsing} shows a graph-based visualization of how our intermediate representation models the scripts shown in Figure~\ref{fig:puppet-script} and Figure~\ref{fig:chef-script}.}
More details about the intermediate representation, including its abstract syntax, can be found in the authors' study on security smells~\cite{saavedra2022glitch}.
We note that minor changes were performed to GLITCH's original intermediate representation: since in technologies such as Ansible, Puppet, and Terraform, we can have nested attributes and variables, we adapted the intermediate representation to consider these nested constructs.\footnote{An updated description of the intermediate representation and its abstract syntax is available in GLITCH's Wiki: \url{https://github.com/sr-lab/GLITCH/wiki/3.-Intermediate-representation}}

\footnotetext{\url{https://github.com/wikimedia/operations-puppet-cdh4/blob/eb24f44669bf6b83fa0ee37bfd1742fb2e297d4c/manifests/hive/metastore/mysql.pp\#L45}}

\subsection{Smell Detection}

To implement a new analysis in GLITCH, developers must extend the abstract class \textit{RuleVisitor}. The \textit{RuleVisitor} class has an abstract method for each component in our intermediate representation. The abstract methods should be implemented to return the list of smells detected on a component. Since the developers have the freedom to implement the abstract methods as they need, they are not bounded to a specific analysis approach. For example, in a previous study focused on security smells~\cite{saavedra2022glitch}, we used a rule-based approach to detect the smells. However, to implement the analyses for the design \& implementation smells, 
we used an algorithmic approach. 

For each Visitor, \GLITCH\ traverses the nodes in the intermediate representation using a depth-first search (DFS). Starting in the initial node (a Project, a Module, or a Unit Block), it executes the DFS considering each collection inside the node as its children. Each node may have more than one code smell, and so every analysis is applied, even if a smell was already identified for that node. \GLITCH\ allows the definition of different configurations to identify code smells. In a Visitor, the developers can define a list of variables, whose values can be changed according to the configuration passed to the tool. Use cases include modifying keywords used for pattern detection or defining a technology-specific token. Configurations allow users to tweak the tool to best suit the needs of the IaC developers and to better adapt to each IaC technology.  If specific behaviour for a technology is required, the Visitors select, in their constructor and according to the technology, objects that check a certain smell, which are called later in the implementation of the methods to check the components of the intermediate representation (see Figure~\ref{fig:specific-behaviour}). 


\subsection{Implementation and Usage}
\GLITCH\ is implemented in Python and it currently supports the analysis of Ansible, Chef, Docker, Puppet, and Terraform scripts. It transforms the original scripts into its intermediate representation and then attempts to detect code smells as described above. To parse Ansible scripts, it uses the \textit{ruamel.yaml} package\footnote{\url{https://pypi.org/project/ruamel.yaml/}} for Python. The Chef scripts are parsed using Ripper,\footnote{\url{https://github.com/ruby/ruby/tree/master/ext/ripper}} a script parser for Ruby. We developed a parser for Ripper's output using a package called \textit{ply}.\footnote{\url{https://github.com/dabeaz/ply}} 
For Docker scripts, we use the packages \textit{dockerfile-parse}\footnote{\url{https://pypi.org/project/dockerfile-parse}} and \textit{bashlex}\footnote{\url{https://pypi.org/project/bashlex}}.
For Puppet scripts, we developed our own parser\footnote{\url{https://github.com/Nfsaavedra/puppetparser}} using the same \textit{ply} package. 
Finally, for Terraform scripts, we use a sligthly modified version of the \textit{python-hcl2} package.\footnote{\url{https://github.com/joaotgoncalves/python-hcl2}}

\unindentedparagraph{\textbf{Using \GLITCH}}
\GLITCH\ provides a command-line interface. Besides the path of the file or folder to analyze, other relevant available options are:
\def\goption#1{\textbf{\texttt{#1}}}
\begin{itemize}
    \item \goption{--tech [ansible|docker|chef|puppet|\\terraform]}: The IaC technology in which the scripts analyzed are written. This option is required.
    \item \goption{--smells [design|security]}: Type of smells being analyzed. Currently it supports nine security smells~\cite{saavedra2022glitch} and \numdesimpsmells\ design \& implementation smells.
    \item \goption{--config PATH}: The path for a config file. Otherwise the default config will be used.
    \item \goption{--tableformat [prettytable|latex]}: The presentation format of the tables that show stats about the analysis.
    \item \goption{--csv}: This flag produces the output in CSV format.
\end{itemize}
The last two options are particularly useful for researchers who need to analyze datasets of IaC scripts and generate CSV data that can be automatically analyzed or tables that can be directly added to research papers. 


\begin{figure}
    \centering
    \scriptsize
    \begin{minted}[]{python}
if tech == Tech.ansible:
    self.imp_align = \
        DesignVisitor.AnsibleImproperAlignmentSmell()
elif tech == Tech.puppet:
    self.imp_align = \
        DesignVisitor.PuppetImproperAlignmentSmell()
(...)
errors += self.imp_align.check(u, u.path)
    \end{minted}
    \caption{Implementation of specific behaviour when creating new rules.}
    \label{fig:specific-behaviour}
\end{figure}


\unindentedparagraph{\textbf{Adding support for new IaC technologies}}
To add a new technology, one needs to create a new subclass of the abstract class \emph{Parser} and to implement the abstract methods that map the constructs of that technology to the intermediate representation (these methods are \textit{parse\_file}, \textit{parse\_folder}, and \textit{parse\_module}).
After that, the developer only needs to change the command-line tool to consider the new technology.

\section{Evaluation}
\label{sec:evaluation}

%
\begin{table}
\centering
\caption{Comparison of \textbf{G}LITCH with \textbf{P}uppeteer \cite{sharma2016does} and \textbf{S}chwarz et al.'s tool \cite{schwarz2018code}. $T_1$/$T_2$ is the percentage of smells detected by $T_1$ that are also detected by $T_2$.}
\label{tab:comparison}
\scriptsize
\begin{tabular}{rcccc}
\toprule
 \textbf{Smells} & \textbf{P/G} (\%) & \textbf{G/P} (\%) & \textbf{S/G} (\%) & \textbf{G/S} (\%) \\
\midrule
Avoid comments & - & - & 100.0 & 100.0 \\
\rowcolor{Gray}
Duplicate block & - & - & 100.0 & 97.6 \\
Improper alignment & 98.4 & 89.9 & 33.3 & 42.9 \\
\rowcolor{Gray}
Long resource & - & - & 82.4 & 82.4 \\
Long statement & 100.0 & 91.4 & 100.0 & 100.0 \\
\rowcolor{Gray}
Misplaced attribute & 100.0 & 100.0 & 97.8 & 97.8 \\
Multifaceted abstraction & - & - & 100.0 & 57.1 \\
\rowcolor{Gray}
Too many variables & - & - & 40.0 & 80.0 \\
Unguarded variable & 100.0 & 92.3 & - & - \\
\midrule
\textbf{Average} & 99.6 & 93.4 & 81.7 & 82.2 \\
\bottomrule
\end{tabular}
\end{table}

\unindentedparagraph{Already conducted studies}
In a previous study focused on security smells for Ansible, Chef, and Puppet~\cite{saavedra2022glitch}, we
used \GLITCH\ to analyze three large datasets containing 196,756 IaC scripts and 12,281,383 LOC. That study demonstrated that \GLITCH\ is robust enough to support a large variety of IaC scripts. It also showed that \GLITCH\ has higher precision and recall than current state-of-the-art tools. 

\unindentedparagraph{Ongoing studies}
We are currently conducting empirical studies similar to Saavedra and Ferreira's study~\cite{saavedra2022glitch}, but focused on the new features implemented in \GLITCH: the new design \& implementation smells and the two new IaC technologies, Docker and Terraform. Due to space limitations, in this paper we focus on presenting preliminary results on the detection of design \& implementation smells for Ansible, Chef, and Puppet.
We use the same three datasets mentioned above; \reviewjffafm{readers interested in the datasets' attributes should consult Table 5 of Saavedra and Ferreira's paper~\cite{saavedra2022glitch}.}



In Table~\ref{tab:comparison}, we compare GLITCH to two state-of-the-art tools that detect design \& implementation smells in Puppet~\cite{sharma2016does} and Chef~\cite{schwarz2018code}. 
\review{We do not compare GLITCH to an Ansible tool since, to the best of our knowledge, GLITCH is the \textbf{first} tool that} \reviewjffafm{detects the design \& implementation smells considered in our study in Ansible scripts.} 
\reviewjffafm{We ran the tools on a subset of the datasets considered.} 
The subset was created by randomly selecting 20 files for each smell, with the smell being detected on each of the selected files. This resulted in a total of 80 Puppet files and 160 Chef files. 
Afterwards, 
we compared the results of \GLITCH\ with the results of the other two tools when considering these files.
We identified smells with the same path, category, and location as reported by each tool.
In some cases, the tools do not output the same line number, although they fundamentally detect the same smell. 
For these cases, we had to manually inspect the files and check whether the tools agree, which was the reason why we only selected a subset of files. 
The replication package has a script that automatically solves the cases we found on this subset of files.

GLITCH can detect almost every smell detected by Puppeteer \cite{sharma2016does} (first column). The value for \emph{Improper alignment} is slightly below 100\% because GLITCH does not consider the alignment in hashes\footnote{\url{https://puppet.com/docs/puppet/latest/lang_data_hash.html}} since these structures, when used as values, are still represented as strings in our intermediate representation. The second column shows that Puppeteer detects a lower percentage of the smells identified by GLITCH. For \emph{Improper alignment}, the reason is that GLITCH, in contrast to Puppeteer, follows the Puppet style guides,\footnote{\url{https://puppet.com/docs/puppet/latest/style_guide.html}} which state that the hash rocket for attributes in a resource should be \textbf{only one space} ahead of the longest attribute name. 

When verifying if GLITCH was able to detect the smells identified by the tool developed by Schwarz et al. \cite{schwarz2018code}, there are two smells with lower percentage values: \textbf{(1)} \emph{Improper alignment} and \textbf{(2)} \emph{Too many variables} (third column). 
This happens because \textbf{(1)} Schwarz et al.'s tool presents false positives for attributes with names such as \textit{variables} and \textit{attributes} because they have structured values which are indented in the lines following the name of the attribute; \textbf{(2)} GLITCH does not consider variable references when calculating the ratio between variables and lines of code. 
Comparing the ability of Schwarz et al.'s tool to detect the smells found by GLITCH (fourth column), there are two smells with a lower percentage: \textbf{(1)} \emph{Improper alignment} and \textbf{(2)} \emph{Multifaceted abstraction}. The main reasons for the lower values are: \reviewjffafm{with respect to \textbf{(1)}, GLITCH detects true positives that are not detected by the other tool and some false positives also undetected by the other tool (the false positives are due to problems when handling blocks, such as conditionals, inside atomic units}); \reviewjffafm{with respect to \textbf{(2)},} GLITCH finds true positives that the other tool does not, since 
Schwarz et al.'s tool does not handle multi-line strings and ignores the pipe character (``$\vert$'').


\section{Related Work}
\label{sec:related}
To the best of our knowledge, \GLITCH\ is the only polyglot code smell detector for IaC scripts.
It unifies other tools such as SLIC~\cite{rahman2019seven}, which detects seven security smells in Puppet scripts, and SLAC~\cite{rahman2021security}, which identifies nine in Chef scripts and six in Ansible scripts (using separate implementations). Compared to SLIC and SLAC, \GLITCH\ identifies two more smells in Puppet scripts and two more in Ansible scripts.
Other relevant tools are \emph{Puppeteer}~\cite{sharma2016does}, which detects design configuration smells, and Schwarz et al.'s tool~\cite{schwarz2018code},
which detects design \& implementation smells.
More recent work includes GASEL~\cite{opdebeeck2023control}, which takes into consideration control-flow and data-flow information for security smell detection in Ansible scripts. GASEL presents better recall and precision than \GLITCH\ for some smells, but it focuses on a single IaC technology compared to five currently supported by \GLITCH.

Finally, some analysis tools for IaC use intermediate representations \cite{ikeshita2017test,shambaugh2016rehearsal,sotiropoulos2020practical} to describe file-system manipulations done by IaC scripts.      
However, to the best of our knowledge, \GLITCH\ is the only smell detector that uses an intermediate representation, allowing a technology-agnostic approach.




\section{Conclusion}
\label{sec:conclusion}
\GLITCH\ has already proven to be a practical and versatile tool. Despite its intermediate representation being remarkably simple, it is expressive enough to enable the implementation of both security and design \& implementation code smells. Furthermore, supporting multiple IaC technologies requires minimal effort, thus reducing the effort of writing code smell analyses for multiple IaC technologies.

Future work includes i) the creation of oracle datasets of design \& implementation smells so that we can measure precision and recall, as we have done in previous studies~\cite{saavedra2022glitch}; ii) the refinement of existing rules and the implementation of new rules to detect more smells; iii) the extension of \GLITCH\ to take control-flow and data-flow information into account, so that we can improve precision and recall~\cite{opdebeeck2023control}.

\section*{Acknowledgments}
Thanks to Akond Rahman for the datasets used in the evaluation of the tools SLIC and SLAC and to Carolina Pereira for her help creating the video demonstrating \GLITCH.
The first author was funded by the Advanced Computing/EuroCC MSc Fellows Programme (EuroHPC grant agreement No 951732). This project was supported by national funds through FCT under project UIDB/50021/2020.

\bibliographystyle{IEEEtran}
\bibliography{references}

\begin{thebibliography}{10}
\providecommand{\url}[1]{#1}
\csname url@samestyle\endcsname
\providecommand{\newblock}{\relax}
\providecommand{\bibinfo}[2]{#2}
\providecommand{\BIBentrySTDinterwordspacing}{\spaceskip=0pt\relax}
\providecommand{\BIBentryALTinterwordstretchfactor}{4}
\providecommand{\BIBentryALTinterwordspacing}{\spaceskip=\fontdimen2\font plus
\BIBentryALTinterwordstretchfactor\fontdimen3\font minus
  \fontdimen4\font\relax}
\providecommand{\BIBforeignlanguage}[2]{{%
\expandafter\ifx\csname l@#1\endcsname\relax
\typeout{** WARNING: IEEEtran.bst: No hyphenation pattern has been}%
\typeout{** loaded for the language `#1'. Using the pattern for}%
\typeout{** the default language instead.}%
\else
\language=\csname l@#1\endcsname
\fi
#2}}
\providecommand{\BIBdecl}{\relax}
\BIBdecl

\bibitem{sharma2016does}
T.~Sharma, M.~Fragkoulis, and D.~Spinellis, ``Does your configuration code
  smell?'' in \emph{2016 IEEE/ACM 13th Working Conference on Mining Software
  Repositories (MSR)}.\hskip 1em plus 0.5em minus 0.4em\relax IEEE, 2016, pp.
  189--200.

\bibitem{schwarz2018code}
J.~Schwarz, A.~Steffens, and H.~Lichter, ``Code smells in infrastructure as
  code,'' in \emph{2018 11th International Conference on the Quality of
  Information and Communications Technology (QUATIC)}.\hskip 1em plus 0.5em
  minus 0.4em\relax IEEE, 2018.

\bibitem{rahman2019seven}
A.~Rahman, C.~Parnin, and L.~Williams, ``The seven sins: Security smells in
  infrastructure as code scripts,'' in \emph{2019 IEEE/ACM 41st International
  Conference on Software Engineering (ICSE)}.\hskip 1em plus 0.5em minus
  0.4em\relax IEEE, 2019, pp. 164--175.

\bibitem{rahman2021security}
A.~Rahman, M.~R. Rahman, C.~Parnin, and L.~Williams, ``Security smells in
  ansible and chef scripts: A replication study,'' \emph{ACM Transactions on
  Software Engineering and Methodology (TOSEM)}, vol.~30, no.~1, 2021.

\bibitem{reis2022leveraging}
S.~Reis, R.~Abreu, M.~d'Amorim, and D.~Fortunato, ``Leveraging practitioners’
  feedback to improve a security linter,'' in \emph{Proceedings of the 37th
  IEEE/ACM International Conference on Automated Software Engineering}, 2022,
  pp. 1--12.

\bibitem{opdebeeck2023control}
R.~Opdebeeck and A.~Zerouali, ``Control and data flow in security smell
  detection for infrastructure as code: Is it worth the effort?'' in
  \emph{Proc. of the 20th Int. Conf. on Mining Software Repositories (MSR
  2023)}, 2023.

\bibitem{saavedra2022glitch}
N.~Saavedra and J.~Ferreira, ``{{GLITCH: Automated Polyglot Security Smell
  Detection in Infrastructure as Code}},'' in \emph{Proceedings of the 37th
  IEEE/ACM International Conference on Automated Software Engineering}, 2022,
  {Preprint available: \url{https://arxiv.org/abs/2205.14371}}.

\bibitem{ikeshita2017test}
K.~Ikeshita, F.~Ishikawa, and S.~Honiden, ``Test suite reduction in idempotence
  testing of infrastructure as code,'' in \emph{International Conference on
  Tests and Proofs}.\hskip 1em plus 0.5em minus 0.4em\relax Springer, 2017, pp.
  98--115.

\bibitem{shambaugh2016rehearsal}
R.~Shambaugh, A.~Weiss, and A.~Guha, ``Rehearsal: A configuration verification
  tool for puppet,'' in \emph{PLDI}, 2016.

\bibitem{sotiropoulos2020practical}
T.~Sotiropoulos, D.~Mitropoulos, and D.~Spinellis, ``{Practical fault detection
  in Puppet programs},'' in \emph{Proceedings of the ACM/IEEE 42nd
  International Conference on Software Engineering}, 2020, pp. 26--37.

\end{thebibliography}
\vspace{12pt}

\end{document}